\documentclass[a4paper]{cas-sc}

\usepackage[authoryear]{natbib}
\usepackage{soul}

\def\tsc#1{\csdef{#1}{\textsc{\lowercase{#1}}\xspace}}
\tsc{WGM}
\tsc{QE}

\usepackage{nomencl}
\usepackage{etoolbox}
\makenomenclature
\usepackage{graphicx}

\renewcommand{\nomgroup}[1]{%
  \item[\bfseries
  \ifstrequal{#1}{A}{Roman symbols}{%
  \ifstrequal{#1}{G}{Greek symbols}{%
  \ifstrequal{#1}{S}{Subscripts}{}}}%
  ]%
}

\begin{document}
\let\WriteBookmarks\relax
\def\floatpagepagefraction{1}
\def\textpagefraction{.001}

\shorttitle{}    

\shortauthors{}  

\title [mode = title]{Numerical analysis of geometrical effect on drop deformation and breakup between rotating cylinders}  



%

\author[1]{Hayato Masuda}[orcid=0000-0003-2870-0423]



\ead{hayato-masuda@omu.ac.jp}


\credit{Conceptualization, Methodology, Software, Validation, Investigation, Data Curation, Writing –Original Draft, Writing – Reviewing\&Editing, Visualization, Project administration, Funding acquisition}

\affiliation[1]{organization={Department of Mechanical Engineering, Graduate School of Engineering, Osaka Metropolitan University},
            addressline={1-1 Gakuen-cho, Naka-ku}, 
            city={Sakai},
            postcode={599-8531}, 
            state={Osaka},
            country={Japan}}

\author[2]{Mitsuhiro Ohta}




\credit{Conceptualization, Methodology, Validation, Writing – Reviewing\&Editing}

\affiliation[2]{organization={Department of Mechanical Science, Graduate school of Technology, Industrial and Social Sciences, Tokushima University},
            addressline={2-1 Minamijyousanjima-cho}, 
            city={Tokushima},
            postcode={7708507}, 
            country={Japan}}

\cortext[1]{Corresponding author}

\author[3]{Tomislav Marić}
\affiliation[3]{organization={Institute of Mathematical Modeling and Analysis, Technische Universität Darmstadt},
            addressline={Peter-Grünberg-Straße 10}, 
            city={Darmstadt},
            postcode={64287}, 
            country={Germany}}

\credit{Software, Validation, Investigation, Data Curation, Writing – Reviewing\&Editing}

\author[3]{Dieter Bothe}
\credit{Investigation, Data Curation, Writing – Reviewing\&Editing}


\begin{abstract}
This study numerically investigates the effects of geometrical curvature on drop deformation and breakup in shear flow between counter-rotating concentric cylinders. Simulations were performed using a geometric Volume-of-Fluid (VOF) method incorporating plicRDF interface reconstruction and isoAdvector interface advection, combined with adaptive mesh refinement. Three circular geometries with different values of the non-dimensional geometrical curvature, $\kappa_{\mathrm F}$, were considered and compared with a planar shear flow system. At fixed Reynolds and Capillary numbers ($Re$ and $Ca$), the breakup morphology differed substantially between the planar and circular systems. In the planar system, the drop underwent relatively symmetric neck thinning, whereas in the circular systems, the drop exhibited asymmetric deformation and pronounced azimuthal elongation. At higher curvature, the drop formed a thin filament that subsequently fragmented, producing multiple satellite drops. Analysis of the internal velocity field and the magnitude of the rate-of-deformation tensor revealed distinct flow and deformation structures associated with these breakup behaviors. Furthermore, the critical $Re$ for breakup, $Re_{\mathrm{cr}}$, decreased with increasing $\kappa_{\mathrm F}$ at nvestigated $Ca$ values. In contrast, the maximum dimensionless elongation length immediately before breakup at $Re=Re_{\mathrm{cr}}$ increased with $\kappa_{\mathrm F}$. These findings suggest that geometrical curvature affects not only the breakup threshold but also the deformation process leading to breakup, providing insights into drop breakup in curved shear flows.
\end{abstract}




\maketitle
\section{Introduction}\label{sec:Introduction}
Drop dynamics under shear flow is important in various processes, such as emulsification \citep{LEstimeetal2024}, liquid–liquid extraction \citep{Sangtametal2021}, liquid–liquid chemical reactions \citep{Manzanoetal2021}, atomization \citep{Daskiranetal2021}, and cooking \citep{Doietal2019}. In these processes, drops are deformed and eventually break up under shear. The breakup behavior mainly depends on the Reynolds and Capillary numbers (\textit{Re} and \textit{Ca}) of the drop. Depending on these parameters, a drop may split into multiple droplets of nearly uniform size or produce numerous small satellite droplets. The droplet size and its distribution after shear processing are critical parameters for process design, as they affect emulsion stability \citep{Goodarzietal2019} and the specific interfacial area \citep{Nagyetal2003}. Therefore, predicting and controlling drop dynamics under shear flow is a fundamental and practically important issue. Despite its fundamental importance, shear-induced drop deformation and breakup continue to attract considerable attention. For example, \citet{Xuetal2020} investigated the effect of shear forces on breakup modes in airflow and experimentally analyzed their complex interactions. \citet{Hardyetal2026} demonstrated that, in turbulent mixing flows, elongational shear near walls promotes drop stretching and induces breakup. As highlighted in these studies, turbulent flow plays an important role in drop dynamics \citep{Ni2024}. However, even in turbulent flows, drops are locally subjected to low or moderate Reynolds number conditions, typically ranging from 0.01 to 100 \citep{Komrakovaetal2014}. Therefore, to understand drop dynamics in this regime, simple shear flow, such as that generated between counter-moving plates, serves as a representative and practically relevant model. So far, numerous experimental studies on drop deformation and breakup under simple shear flow have been conducted, e.g., \citeauthor{Taylor1934} (\citeyear{Taylor1934}); \citeauthor{Stoneetal1986} (\citeyear{Stoneetal1986}); \citeauthor{Sibilloal2006} (\citeyear{Sibilloal2006}); \citeauthor{Lietal2010} (\citeyear{Lietal2010}). However, for this purpose, numerical simulations are a more powerful tool, as they enable detailed evaluation of local information such as interfacial structure and stress distribution.

As pioneering numerical studies, \citet{Lietal2000} reported drop deformation and breakup in shear flow between parallel plates. They employed the volume-of-fluid (VOF) method to capture and track the drop interface and successfully demonstrated a sequence of pinch-off phenomena. Subsequently, \citet{Renardyetal2001} investigated the effect of inertia on drop breakup in more detail, varying \textit{Re} over a wider range than in their previous work \citep{Lietal2000}. As a result, they derived the critical condition for drop breakup as
\begin{equation}
    Re \sim \frac{1}{Ca}.
\end{equation}
This relationship implies that $Re \cdot Ca$ remains approximately constant at the critical condition, indicating that the critical \textit{Ca} (or \textit{Re}) decreases with increasing \textit{Re} (or \textit{Ca}). Moreover, \citet{Renardyetal2002} numerically reproduced droplet fragmentation at higher \textit{Re} using a piecewise linear interface calculation (PLIC) method for interface reconstruction. Despite these pioneering efforts (\citeauthor{Lietal2000}, \citeyear{Lietal2000}; \citeauthor{Renardyetal2001}, \citeyear{Renardyetal2001}; \citeauthor{Renardyetal2002}, \citeyear{Renardyetal2002}), numerical studies of droplet deformation under shear flow, despite the apparent simplicity of the system, remain an active area of research. For example, \citet{Komrakovaetal2015} examined the effect of viscosity ratio (ratio of dispersed- to continuous-phase viscosities) on drop breakup using the lattice Boltzmann method and showed that the critical \textit{Ca} decreases with increasing viscosity ratio. \citet{Hernandezetal2017} investigated drop breakup in highly confined geometries and found that droplets become more stable as confinement increases due to wall effects. \citet{Lietal2024} analyzed the effects of drop diameter, viscosity ratio, and wall confinement, in addition to \textit{Re} and \textit{Ca}, and discussed breakup conditions based on the balance between dynamic pressure and interfacial and viscous stresses. More recently, \citet{Ohtaetal2025} reported approximate critical conditions for drop and bubble breakup in terms of \textit{Re} and \textit{Ca} using a coupled level set and VOF (CLSVOF) method. Notably, their study covered low-\textit{Ca} and high-\textit{Re} conditions for which experimental and numerical data are scarce. As indicated by this literature survey, the critical conditions for drop breakup, which are essential for predicting droplet size, appear to have been well established for simple shear flow.

However, previous studies have primarily focused on drops in planar shear flows, where the streamlines have no curvature. In practice, however, shear flows are often curved, and thus understanding drop deformation and breakup in curved shear fields is also important. For instance, \citet{Dressleretal2008} numerically investigated droplet deformation and breakup in immiscible polymer blends between concentrically rotating cylinders, considering different cylinder rotation conditions and gap widths. Their study focused on the deformation and breakup characteristics of droplets in viscoelastic polymer blends, particularly in relation to the production of small droplets. Nevertheless, the influence of geometrical curvature on the deformation and critical breakup conditions of an individual Newtonian drop remains insufficiently understood. This gap is important because critical breakup conditions have mostly been discussed based on planar shear flow, whereas the role of streamline curvature in shear-induced drop breakup remains unclear. Here, a fundamental question arises: how do drops behave in shear flow between rotating cylinders? As noted by \citet{Janssenetal1994}, such a flow can be approximated as planar shear when the radius of curvature is sufficiently large. Therefore, investigating drop dynamics at small to moderate radii of curvature is of practical importance.

In this study, we numerically investigate drop dynamics in shear flow generated between counter-rotating cylinders with three different geometries (see the next section). In particular, the critical conditions for drop breakup are evaluated using the VOF method, and the effects of geometry are examined through comparison with the planar shear case.

\section{Numerical analysis}\label{sec:Numerical analysis}
\subsection{Computational domain}\label{subsec:Computational domain}
\begin{figure}[pos=!htbp]
    \centering
    \includegraphics[width=0.7\linewidth]{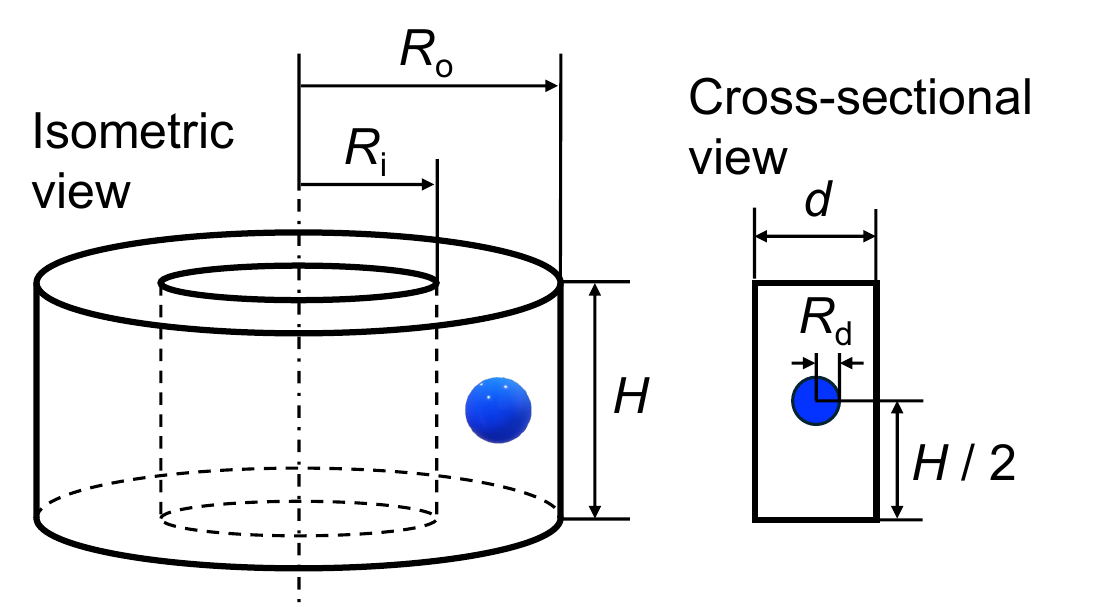}
    \caption{Computational domain for circular shear flow system.}
    \label{fig1}
\end{figure}

\begin{figure}[pos=!htbp]
    \centering
    \includegraphics[width=0.5\linewidth]{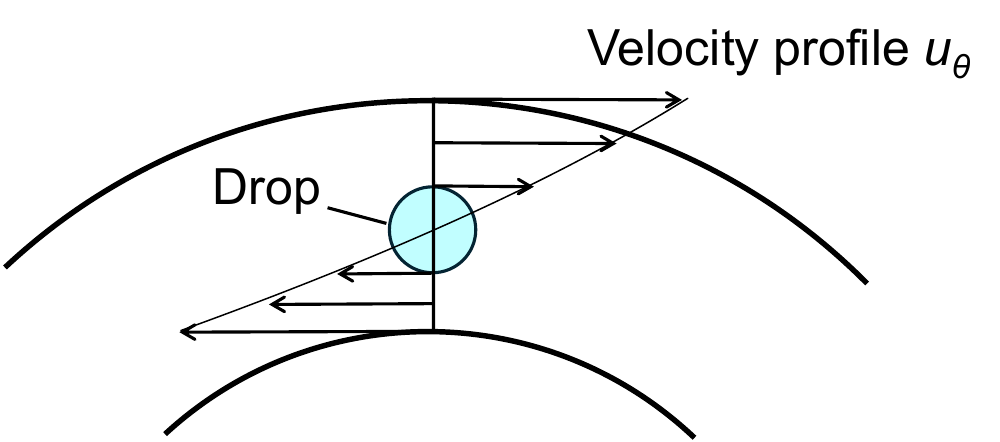}
    \caption{Initial azimuthal velocity profile in circular shear flow system.}
    \label{fig2}
\end{figure}

\textbf{Fig.~\ref{fig1}} illustrates the computational domain consisting of two concentric cylinders. Three geometrical configurations with different inner and outer cylinder radii ($R_{\mathrm i}$ and $R_{\mathrm o}$) were examined: ($R_{\mathrm i}$, $R_{\mathrm o}$) = (0.025 m, 0.038 m), (0.018 m, 0.031 m), and (0.0125 m, 0.0255 m), hereafter referred to as Types 1, 2, and 3, respectively. The gap width, $d$, and axial height, $H$, were fixed at 0.013 m and 0.026 m, respectively, for all configurations. The non-dimensional geometrical curvature, $\kappa_{\mathrm F}$, was defined as $\kappa_{\mathrm F}=d/R_{\mathrm i}$, yielding $\kappa_{\mathrm F}$ = 0.52, 0.72, and 1.04 for Types 1, 2, and 3, respectively. Periodic boundary conditions were imposed on the top and bottom boundaries. A shear flow was generated by rotating the inner and outer cylinders in opposite directions at angular velocities $\omega_{\mathrm i}$ and $\omega_{\mathrm o}$, respectively. The inner and outer cylinders were rotated in opposite directions with equal magnitudes of azimuthal wall velocity, i.e., $|R_{\mathrm i}\omega_{\mathrm i}|=|R_{\mathrm o}\omega_{\mathrm o}|$. The theoretical azimuthal velocity profile between the rotating cylinders is given by
\begin{equation}
u_{\theta}=\left(\frac{\omega_{\mathrm o}R_{\mathrm o}^2-\omega_{\mathrm i}R_{\mathrm i}^2}{R_{\mathrm o}^2-R_{\mathrm i}^2}\right)r+\frac{(\omega_{\mathrm i}-\omega_{\mathrm o})R_{\mathrm i}^2R_{\mathrm o}^2}{R_{\mathrm o}^2-R_{\mathrm i}^2}\frac{1}{r},
\label{eq:vpro}
\end{equation}
where $r$ is the radial coordinate. The drop, with an initial radius of $R_{\mathrm d}=0.0025$ m, was initially centered at the radial position $r_0$, where the azimuthal velocity of the corresponding single-phase Couette flow vanishes, i.e., $u_\theta(r_0)=0$, as determined from Eq.~\ref{eq:vpro} and illustrated in \textbf{Fig.~\ref{fig2}}. The geometrical parameters and initial drop positions for each configuration are summarized in Table~\ref{tab:geometry}.
\begin{table}[htbp]
\centering
\caption{Geometrical parameters and initial drop positions for the circular systems.}
\label{tab:geometry}
\begin{tabular}{lcccc}
\toprule
Geometry & $R_i$ [m] & $R_o$ [m] & $\kappa_{\mathrm F}$ & $r_0$ [m]\\
\midrule
Type 1 & 0.025 & 0.038 & 0.52 & 0.03082\\
Type 2 & 0.018 & 0.031 & 0.72 & 0.02362\\
Type 3 & 0.0125 & 0.0255 & 1.04 & 0.01785\\
\bottomrule
\end{tabular}
\end{table}

Drop dynamics was characterized using the Reynolds and Capillary numbers, $Re$ and $Ca$, defined based on the initial drop radius as
\begin{equation}
Re=\frac{\rho_{\mathrm s}\dot{\gamma}_{\mathrm R}R_{\mathrm d}^2}{\mu_{\mathrm s}},
\label{eq:Re}
\end{equation}
\begin{equation}
Ca=\frac{\mu_{\mathrm s}\dot{\gamma}_{\mathrm R}R_{\mathrm d}}{\sigma},
\label{eq:Ca}
\end{equation}
where $\rho_{\mathrm s}$ and $\mu_{\mathrm s}$ are the density and dynamic viscosity of the surrounding fluid, respectively, $\sigma$ is the interfacial tension, and $\dot{\gamma}_{\mathrm R}$ is the representative shear-rate. In the circular geometry, the local shear-rate varies with radial position and is expressed as $\dot{\gamma}(r)=\left|du_{\theta}/dr-u_{\theta}/r\right|$. The shear-rate at the initial drop center, $r_0$, was selected as the representative shear-rate, $\dot{\gamma}_{\mathrm R}=\dot{\gamma}(r_0)$. Since $u_\theta(r_0)=0$, this can be expressed as
\begin{equation}
\dot{\gamma}_{\mathrm R}=\left|\frac{\omega_{\mathrm o}R_{\mathrm o}^2-\omega_{\mathrm i}R_{\mathrm i}^2}{R_{\mathrm o}^2-R_{\mathrm i}^2}-\frac{(\omega_{\mathrm i}-\omega_{\mathrm o})R_{\mathrm i}^2R_{\mathrm o}^2}{R_{\mathrm o}^2-R_{\mathrm i}^2}\frac{1}{r_0^2}\right|.
\label{eq:gammaCircular}
\end{equation}
$Re$ was varied by adjusting the angular velocities of the inner and outer cylinders, $\omega_{\mathrm i}$ and $\omega_{\mathrm o}$, while maintaining equal magnitudes of their azimuthal wall velocities. $Ca$ was independently controlled by adjusting the interfacial tension, $\sigma$, for each $Re$. Note that no Taylor vortices were observed under the present computational conditions.

\subsection{Mathematical formulation and numerical methods}
\label{subsec:Mathematical formulation and numerical methods}
\begin{figure}[pos=!htbp]
    \centering
    \includegraphics[width=0.3\linewidth]{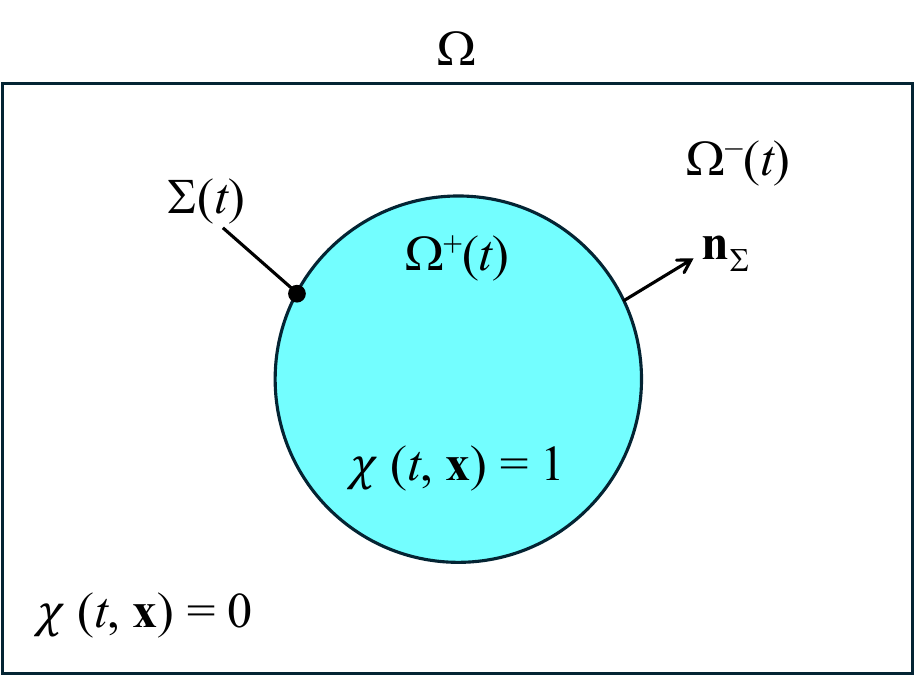}
    \caption{Schematic illustration of two-phase computational domain, $\Omega$.}
    \label{fig3}
\end{figure}

\textbf{Fig.~\ref{fig3}} illustrates a typical system of incompressible, immiscible two-phase flow without phase change. The computational domain, $\Omega$, is divided by the interface $\Sigma(t)$ into two subdomains, $\Omega^+(t)$ and $\Omega^-(t)$, occupied by the two fluids. Here, $\mathbf n_{\Sigma}$ denotes the unit normal vector at the interface pointing outward from $\Omega^+(t)$.

A phase indicator function is introduced to distinguish the two phases:
\begin{equation}
\chi(\mathbf{x},t):=
\begin{cases}
1, & \mathbf{x}\in\Omega^+(t),\\
0, & \mathbf{x}\in\Omega^-(t).
\end{cases}
\end{equation}
Using the phase indicator function $\chi$, the volume fraction of the phase occupying $\Omega^+(t)$ within a fixed control volume $\Omega_{\mathrm c}$, denoted as $\alpha_{\mathrm c}$, is defined as
\begin{equation}
\alpha_{\mathrm c}(t):=\frac{1}{|\Omega_{\mathrm c}|}\int_{\Omega_{\mathrm c}}\chi(\mathbf{x},t)\,dV.
\end{equation}
In the VOF method, $\alpha_{\mathrm c}$ indicates the fraction of each computational cell occupied by the phase $\Omega^+(t)$ and is classified as
\begin{equation}
\alpha_{\mathrm c}=
\begin{cases}
1, & \text{cell is entirely occupied by } \Omega^+(t),\\
\in(0,1), & \text{cell contains the interface},\\
0, & \text{cell is entirely occupied by } \Omega^-(t).
\end{cases}
\end{equation}

The local density and viscosity are expressed in terms of $\alpha_{\mathrm c}$. Assigning $\Omega^+(t)$ and $\Omega^-(t)$ to the drop and surrounding fluid, respectively, these properties are calculated as
\begin{equation}
\rho=\alpha_{\mathrm c}\rho^++(1-\alpha_{\mathrm c})\rho^-=\alpha_{\mathrm c}\rho_{\mathrm d}+(1-\alpha_{\mathrm c})\rho_{\mathrm s},
\label{mixture density}
\end{equation}
\begin{equation}
\mu=\alpha_{\mathrm c}\mu^++(1-\alpha_{\mathrm c})\mu^-=\alpha_{\mathrm c}\mu_{\mathrm d}+(1-\alpha_{\mathrm c})\mu_{\mathrm s},
\label{mixture viscosity}
\end{equation}
where the subscripts $\mathrm d$ and $\mathrm s$ denote the drop and surrounding fluid, respectively. Although drop dynamics can be affected by the density and viscosity ratios between the two fluids \citep{Komrakovaetal2015,Lietal2024}, both ratios were set to unity in the present study for simplicity, i.e., $\rho_{\mathrm d}=\rho_{\mathrm s}$ and $\mu_{\mathrm d}=\mu_{\mathrm s}$.

The mass and momentum conservation equations for incompressible two-phase flow in the single-field formulation are given by
\begin{equation}
\nabla\cdot\mathbf u=0,
\label{eq:continuity}
\end{equation}
\begin{equation}
\frac{\partial(\rho\mathbf u)}{\partial t}+\nabla\cdot(\rho\mathbf u\otimes\mathbf u)=-\nabla p-(\mathbf g\cdot{\mathbf x})\nabla\rho+\nabla\cdot(2\mu\mathbf D)+\rho\mathbf g+\mathbf f_{\Sigma},
\label{eq:NS}
\end{equation}
where $\mathbf u$ is the velocity vector, $p$ is the modified pressure defined as the subtraction of hydrostatic pressure from the total pressure \citep{Liuetal2024}, $\mathbf D=[\nabla\mathbf u+(\nabla\mathbf u)^T]/2$ is the rate-of-deformation tensor, and $\mathbf g$ is the gravitational acceleration. The surface tension force, $\mathbf f_{\Sigma}$, is represented using the continuum surface force (CSF) formulation \citep{Brackbilletal1992} as
\begin{equation}
\mathbf f_{\Sigma}:=\sigma\kappa\mathbf n_{\Sigma}\delta_{\Sigma},
\end{equation}
where $\sigma$ is the interfacial tension, $\kappa$ is the interfacial curvature, and $\delta_{\Sigma}$ is the Dirac delta distribution concentrated at the interface $\Sigma(t)$.

The transport equation for the phase indicator function is integrated over a fixed control volume $\Omega_{\mathrm c}$ during a time interval $[t^n,t^{n+1}]$. Applying the Reynolds transport theorem, the evolution of the cell-averaged volume fraction can be expressed in the following integral form \citep{Maricetal2020,Asgharetal2025}:
\begin{equation}
\alpha_{\mathrm c}(t^{n+1})=\alpha_{\mathrm c}(t^n)-\frac{1}{|\Omega_{\mathrm c}|}\int_{t^n}^{t^{n+1}}\int_{\partial\Omega_{\mathrm c}}\chi(t,\mathbf{x})\,\mathbf u\cdot\mathbf n\,dS\,dt,
\label{eq:volumeFractionTransportEq}
\end{equation}
where $\partial\Omega_{\mathrm c}$ denotes the boundary of the control volume and $\mathbf n$ is the outward unit normal vector to the boundary. Since the cell boundary consists of individual faces, the surface integral can be decomposed into a sum of integrals over the cell faces. Defining $V_f^\alpha$ as the signed phase-specific volume transported across each face $S_f$ during the time interval $[t^n,t^{n+1}]$, the volume fraction transport equation can be written as
\begin{equation}
\alpha_{\mathrm c}(t^{n+1})=\alpha_{\mathrm c}(t^n)-\frac{1}{|\Omega_{\mathrm c}|}\sum_{f\in F_c}V_f^\alpha,
\end{equation}
where $F_c$ denotes the set of faces of the control volume. This formulation provides the basis for geometric VOF methods, which differ in how the transported phase-specific volume $V_f^\alpha$ is evaluated \citep{Maricetal2020}.

Among the available geometric VOF methods, the plicRDF--isoAdvector method was employed in this study \citep{Scheufleretal2019}. This method combines plicRDF interface reconstruction with isoAdvector interface advection. In the plicRDF method, the interface is reconstructed using reconstructed distance functions (RDFs) evaluated in the neighborhood of interfacial cells. The volume fraction is then transported using the isoAdvector method \citep{Roenbyetal2016}. In this approach, the phase-specific flux across each cell face is evaluated based on the reconstructed interface geometry. The transported volume is determined by integrating the time-dependent area of the cell face occupied by the corresponding phase. The resulting fluxes are used to update the volume fraction within the finite-volume framework. Numerical simulations were performed using the interFlow solver from our fork of the TwoPhaseFlow project \citep{TwoPhaseFlow}, implemented in OpenFOAM v2312. Second-order discretization schemes were employed for the numerical simulations.

\subsection{Validation of computational code}
\label{subsec:Validation of computational code}
\begin{figure}[pos=!htbp]
    \centering
    \includegraphics[width=0.5\linewidth]{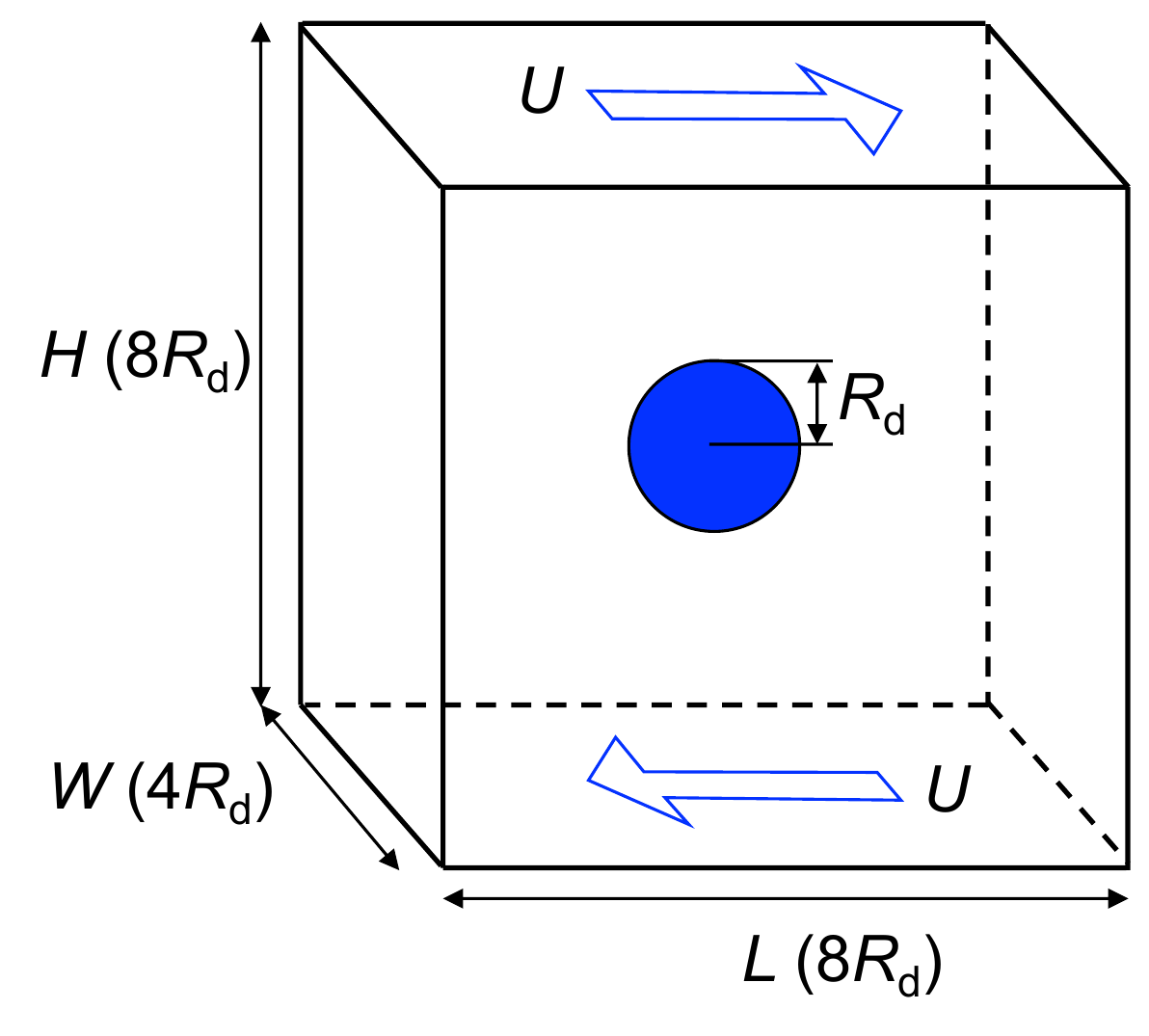}
    \caption{Computational domain for planar shear flow system.}
    \label{fig4}
\end{figure}

\begin{figure}[pos=!htbp]
    \centering
    \includegraphics[width=0.8\linewidth]{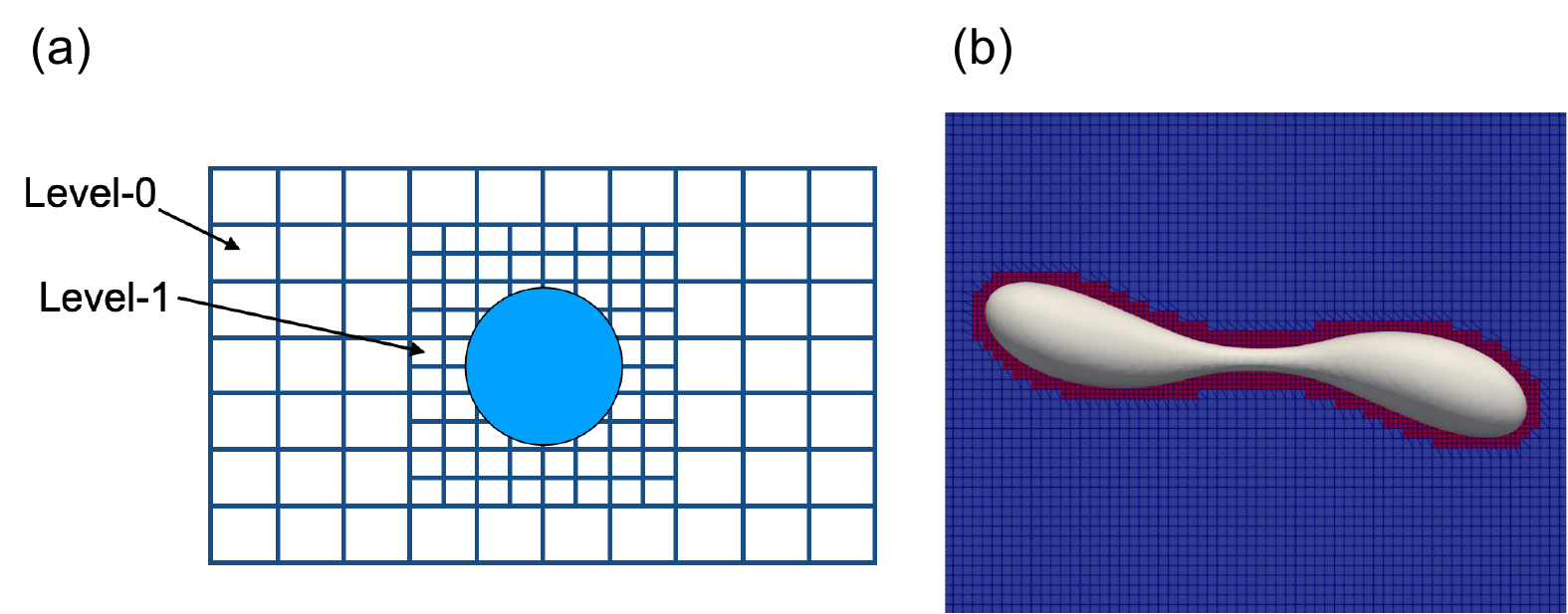}
    \caption{Two-level adaptive mesh refinement (AMR): (a) schematic illustration of initial configuration and (b) snapshot of deforming drop with refined meshes.}
    \label{fig5}
\end{figure}

To validate the numerical approach, simulations of drop deformation and breakup in planar shear flow were performed, and the results were compared with those reported by \citet{Lietal2000} using the VOF method and \citet{Ohtaetal2025} using the coupled level-set and volume-of-fluid (CLSVOF) method. The computational domain, with dimensions of $L=8R_{\mathrm d}$, $W=4R_{\mathrm d}$, and $H=8R_{\mathrm d}$, is illustrated in \textbf{Fig. ~\ref{fig4}}. This configuration is identical to those used by \citet{Lietal2000} and \citet{Ohtaetal2025}. The upper and lower plates moved in opposite directions with velocities of $U$ and $-U$, respectively. This configuration generates a linear velocity profile with a constant shear-rate of $\dot{\gamma}_{\mathrm R}=2U/H$. Accordingly, $Re$ and $Ca$ for the planar geometry were calculated as
\begin{equation}
Re=\frac{\rho_{\mathrm s}(2U/H)R_{\mathrm d}^2}{\mu_{\mathrm s}},
\label{eq:Rep}
\end{equation}
\begin{equation}
Ca=\frac{\mu_{\mathrm s}(2U/H)R_{\mathrm d}}{\sigma},
\label{eq:Cap}
\end{equation}
The density and viscosity ratios between the drop and surrounding fluid were both set to unity, consistent with the circular shear flow simulations. As illustrated in \textbf{Fig. ~\ref{fig5}}, two-level adaptive mesh refinement (AMR) was employed to resolve the drop interface with finer cells. The base and refined mesh levels were denoted by $\ell=0$ and $\ell=1$, respectively. \citet{Lietal2000} performed simulations at $Ca=0.3$ and various $Re$ using a uniform mesh with $\Delta x=\Delta y=\Delta z=R_{\mathrm d}/16$. Subsequently, \citet{Ohtaetal2025} validated their CLSVOF method against the results of \citet{Lietal2000} using two-level AMR, with cell sizes of $\Delta x^{\ell=0}=\Delta y^{\ell=0}=\Delta z^{\ell=0}=R_{\mathrm d}/8$ and $\Delta x^{\ell=1}=\Delta y^{\ell=1}=\Delta z^{\ell=1}=R_{\mathrm d}/16$. The same mesh resolutions were adopted in the present validation simulations.

\begin{figure}[pos=!htbp]
    \centering
    \includegraphics[width=0.4\linewidth]{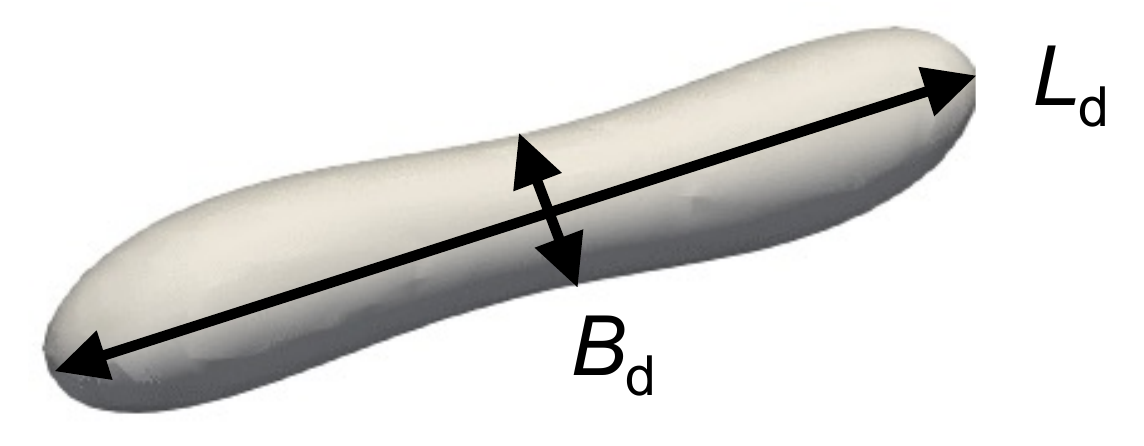}
    \caption{Snapshot of a deformed drop in planar shear flow at $Re=0.75$ and $Ca=0.3$.}
    \label{fig6}
\end{figure}

\textbf{Fig. ~\ref{fig6}} illustrates the deformed drop at $Re=0.75$ and $Ca=0.3$, indicating its major and minor axes. To quantify drop deformation, the Taylor deformation parameter, $D$, was defined as
\begin{equation}
D=\frac{L_{\mathrm d}-B_{\mathrm d}}{L_{\mathrm d}+B_{\mathrm d}},
\label{eq:D}
\end{equation}
where $L_{\mathrm d}$ and $B_{\mathrm d}$ denote the major and minor axes of the deformed drop, respectively. The parameter $D$ characterizes the overall drop deformation, with $D=0$ corresponding to an undeformed spherical drop, while $D$ approaches unity as the drop becomes highly elongated and approaches breakup. The values of $D$ obtained at different $Re$ are compared with published results in Table~\ref{tab:comparisonD}. The present predictions showed good agreement with the deformation parameters reported by \citet{Lietal2000} and \citet{Ohtaetal2025}. Furthermore, breakup was consistently predicted at $Re=0.75$, in agreement with both previous studies. These results demonstrate that the present numerical approach reproduces the reported drop deformation and breakup behavior under the investigated conditions.

\begin{table}[htbp]
\centering
\caption{Comparison of the Taylor deformation parameter, $D$, obtained in the present study and previous studies.}
\label{tab:comparisonD}
\begin{tabular}{lccc}
\toprule
 & This study & \citet{Lietal2000} & \citet{Ohtaetal2025}\\
\midrule
$Re = 0.1$  & $0.389$  & $0.3968$ & $0.3960$\\
$Re = 0.5$  & $0.450$  & $0.453$ & $0.4570$\\
$Re = 0.6$  & $0.4721$ & $0.4768$ & $0.458$\\
$Re = 0.75$ & Breakup & Breakup & Breakup\\
\bottomrule
\end{tabular}
\end{table}

\subsection{Examination of mesh resolution}
\label{subsec:Examination of mesh resolution}
The effect of mesh resolution on drop deformation and breakup in the circular system was examined using the Type 1 geometry ($R_{\mathrm i}=0.025$ m, $R_{\mathrm o}=0.038$ m, and $\kappa_{\mathrm F}=0.52$) at $Re=1.0$ and $Ca=0.3$. Three base mesh resolutions were considered: $\Delta x^{\ell=0}\approx R_{\mathrm d}/4$, $R_{\mathrm d}/8$, and $R_{\mathrm d}/10$. As described in Section~\ref{subsec:Validation of computational code}, two-level adaptive mesh refinement (AMR) was employed inside the drop and near the interface, resulting in finest-cell sizes of $\Delta x^{\ell=1}\approx R_{\mathrm d}/8$, $R_{\mathrm d}/16$, and $R_{\mathrm d}/20$, respectively. The corresponding initial numbers of computational cells before AMR were 252,800, 2,022,400, and 4,283,136. The total number of cells increased during the simulations owing to AMR. Hereafter, the mesh resolution is represented by the finest-cell size, $\Delta x^{\ell=1}$.

\begin{figure}[pos=!htbp]
    \centering
    \includegraphics[width=0.5\linewidth]{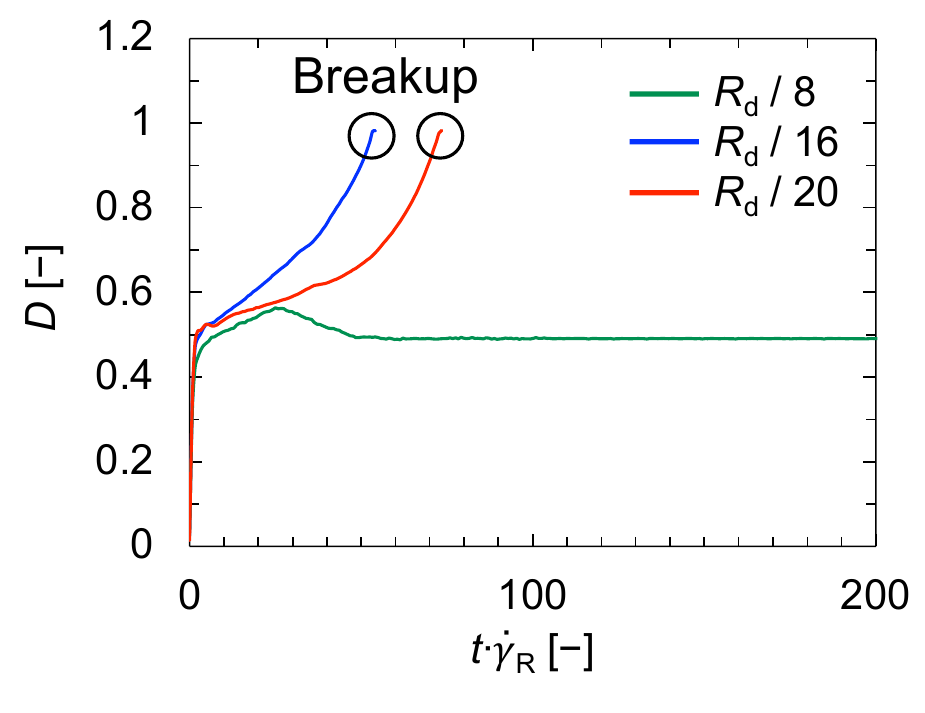}
    \caption{Time evolution of Taylor deformation parameter, $D$, as a function of dimensionless time, $t\cdot\dot{\gamma}_{\mathrm R}$, at $Re$ = 1.0 and $Ca$ = 0.3 for different mesh resolutions.}
    \label{fig7}
\end{figure}

\textbf{Fig.~\ref{fig7}} shows the temporal evolution of the Taylor deformation parameter, $D$, as a function of the dimensionless time, $t\dot{\gamma}_{\mathrm R}$. The initial deformation behavior was similar among the three mesh resolutions. However, clear differences emerged as deformation progressed. For the coarsest mesh ($R_{\mathrm d}/8$), the deformation parameter approached a nearly constant value without breakup during the simulation. In contrast, $D$ continued to increase for the two finer meshes ($R_{\mathrm d}/16$ and $R_{\mathrm d}/20$), eventually leading to breakup at $t\cdot\dot{\gamma}_{\mathrm R}=54.12$ and $73.50$, respectively. Although both finer meshes predicted drop breakup, the differences in their deformation histories and breakup times indicate that the quantitative predictions remain sensitive to mesh resolution. Nevertheless, previous VOF-based studies of drop and bubble breakup have adopted comparable mesh resolutions of approximately $R/16$ \citep{Kekesietal2014,Ohtaetal2025}. Considering the qualitative agreement in breakup prediction between the two finer meshes in the present study and the computational cost, a finest-cell size of approximately $R_{\mathrm d}/16$ was selected for the subsequent circular shear flow simulations.

\section{Results and discussion}\label{sec:Results and discussion}
\subsection{Geometrical effect on drop dynamics}\label{subsec:Geometrical effect on drop dynamics}
The drop dynamics was compared between planar and circular systems (Type 1-3). This comparison covered the range of $\kappa_{\rm F}$ from $0$ to $1.04$. Noted that $\kappa_{\mathrm F} = 0$ corresponds to the planar system. Unlike the validation case, in which the domain size was chosen to match the previous studies (\citeauthor{Lietal2000}, \citeyear{Lietal2000}; \citeauthor{Ohtaetal2025}, \citeyear{Ohtaetal2025}), the computational domain for the present simulations was designed to enable a direct comparison between the planar and circular geometries. This treatment is important because drop dynamics can be significantly affected by the geometrical configuration, even in planar shear flows, including the gapwise distance between the counter-moving walls and the spanwise extent of the domain (\citeauthor{Sibilloal2006}, \citeyear{Sibilloal2006}; \citeauthor{Hernandezetal2017}, \citeyear{Hernandezetal2017}; \citeauthor{Lietal2024}, \citeyear{Lietal2024}; \citeauthor{Ohtaetal2025}, \citeyear{Ohtaetal2025}). In the planar system, $L$, $W$, and $H$ represent the streamwise, spanwise, and gapwise directions, respectively, with the upper and lower walls moving in opposite directions along $L$. In the circular system, these directions correspond to the circumferential, axial, and radial directions, respectively. Since the circular system used $R_\mathrm{d}=0.0025\ \mathrm{m}$, a gap width of $d=0.013\,\mathrm{m}$, and a cylinder height of $0.026\,\mathrm{m}$, the planar dimensions $W$ and $H$ were set to $10.4R_\mathrm{d}$ and $5.2R_\mathrm{d}$, respectively. This setting was based on previous reports (\citeauthor{Ohtaetal2025}, \citeyear{Ohtaetal2025}) indicating that the critical Reynolds number for breakup becomes nearly independent of $L$ when $L\geqq24R_{\mathrm d}$. Thus, $L=26R_{\mathrm d}$ was adopted to minimize possible effects of the streamwise domain length. Consequently, the planar domain was set to $L(10.4R_\mathrm{d})\times W(10.4R_\mathrm{d})\times H(5.2R_\mathrm{d})$.

\begin{figure}[pos=!htbp]
    \centering
    \includegraphics[width=1.0\linewidth]{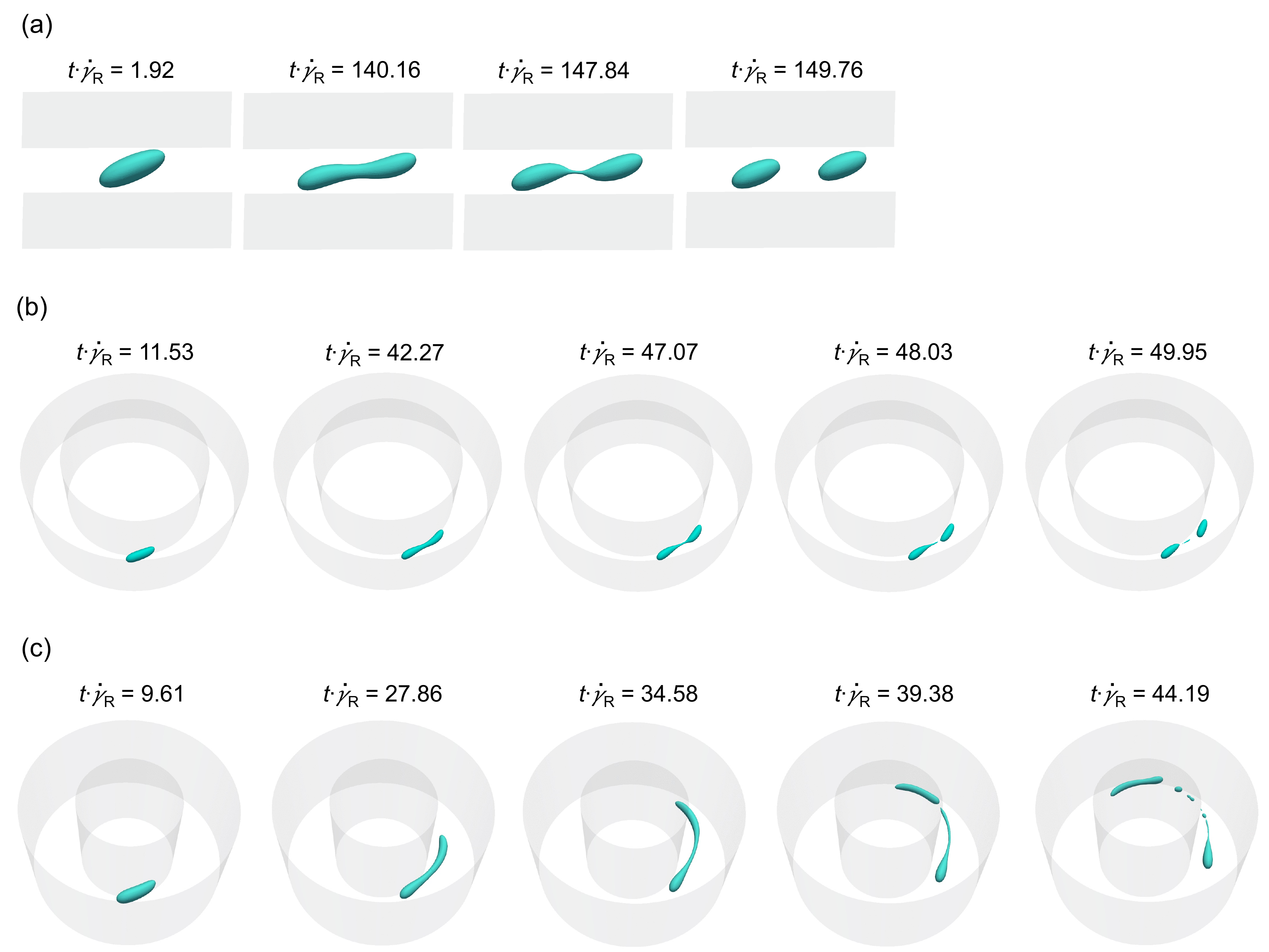}
    \caption{Drop deformation and breakup at $Re$ = 1.2 and $Ca$ = 0.3: (a) planar system, (b) Type 1, and (c) Type 3.}
    \label{fig8}
\end{figure}

\textbf{Fig. ~\ref{fig8}} shows the drop dynamics for each geometry (planar system, Types 1 and 3) at $Re = 1.2$ and $Ca = 0.3$, with a focus on the breakup process. Under this condition, the drop eventually broke up in all geometries. However, the breakup morphology depended strongly on the geometry, despite the same hydrodynamic non-dimensional parameters, namely $Re$ and $Ca$.

In the planar system, the drop underwent a typical pinch-off process, resulting in a nearly symmetric breakup into daughter drops of comparable size. This behavior is consistent with the breakup dynamics reported in previous studies for drops in shear flows, such as those by \citet{Lietal2000} and \citet{Ohtaetal2025}. In this case, the drop first deformed into an inclined prolate shape and subsequently formed a neck near the central region of the elongated drop. Although the drop was tilted by the shear flow, the necking process remained relatively balanced, leading to a nearly binary breakup.
In contrast, the breakup process in the circular geometries was accompanied by azimuthal migration of the drop. In these cases, the drop centroid was initially placed at the radial position where the azimuthal velocity of the corresponding single-phase Couette flow becomes zero. However, because the azimuthal velocity distribution given by Eq.~\ref{eq:vpro} is nonlinear in the radial direction, the inner- and outer-cylinder sides of the finite-size drop experience different local velocities. Although the magnitudes of the tangential velocities at the inner and outer cylinder surfaces, $\omega_i R_i$ and $\omega_o R_o$, were set to be identical, the nonlinear azimuthal velocity profile given by Eq.~\ref{eq:vpro} causes a velocity imbalance across the finite-size drop. In particular, the inner-cylinder side of the drop experiences a larger azimuthal velocity than the outer-cylinder side. As a result, the drop does not remain at the initial zero-velocity position but migrates in the direction of the inner-cylinder rotation. In Type 1, this curvature-induced migration was accompanied by a slightly asymmetric deformation. A neck-like structure was formed prior to breakup, as in the planar case. However, the two sides of the neck did not elongate equally, and the resulting daughter drops were no longer perfectly symmetric. At $Re = 1.2$ and $Ca = 0.3$, the breakup was accompanied by a small satellite drop. This indicates that even weak curvature modifies the fore--aft symmetry of the pinch-off process and affects the final stage of drop breakup. With increasing curvature, this tendency became more pronounced. In Type 3, the breakup pathway was markedly different from that in the planar system. The drop was strongly advected and stretched along the azimuthal direction of the cylindrical geometry, forming a long, flattened, and crescent-like morphology around the inner cylinder. Unlike the planar case, where the deformation was characterized by localized necking near the central region, the Type 3 drop developed a thin filament-like region as a result of global azimuthal elongation. This elongated region became progressively thinner and eventually ruptured into multiple fragments, accompanied by several satellite drops. A similar filament-mediated breakup accompanied by strong elongation and satellite drops has also been reported in planar shear flows at higher $Re$, where inertial effects become more pronounced \citep{Renardy2001,Jeong2026}. It should be noted that such high-$Re$ breakup behavior in planar shear flows can also become asymmetric. In the present Type 3 case, however, a strongly elongated and asymmetric breakup morphology appeared under the same nominal $Re$ and $Ca$ as the planar case. This indicates that curvature promotes azimuthal elongation and asymmetric fragmentation through curvature-induced azimuthal migration and fore--aft asymmetry of the deforming drop, rather than through an increase in the nominal inertia alone.

These results suggest that the effect of curvature is not merely to shift the critical condition for breakup. Rather, curvature alters the entire breakup pathway. Even at the same nominal $Re$ and $Ca$, the breakup pattern changes qualitatively with curvature: from a nearly symmetric binary pinch-off in the planar system, to weakly asymmetric breakup with a satellite drop in Type 1, and finally to highly elongated, filament-mediated fragmentation with multiple satellite drops in Type 3.

\begin{figure}[pos=!htbp]
    \centering
    \includegraphics[width=0.85\linewidth]{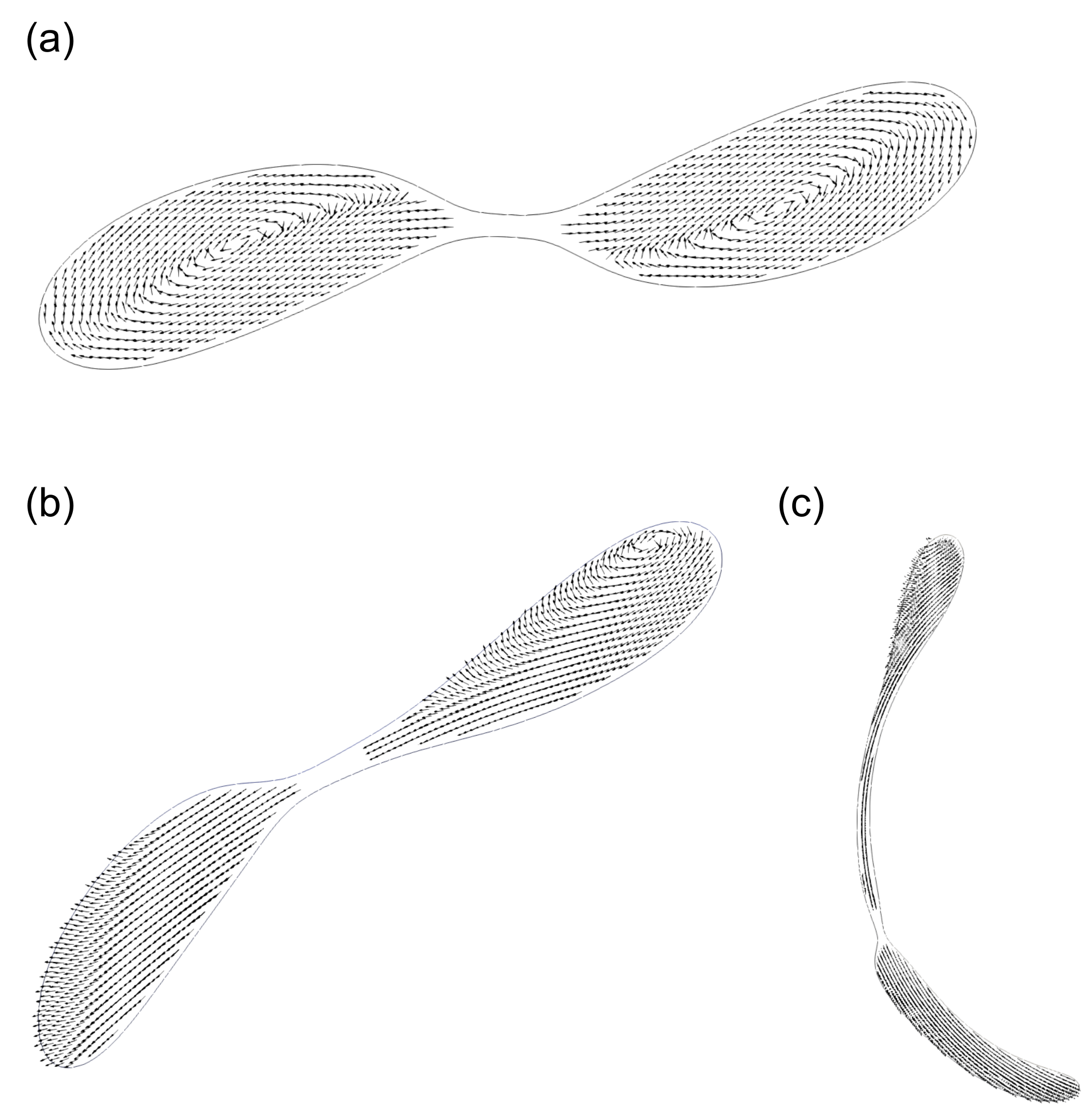}
    \caption{Velocity vectors inside drop relative to its translational motion immediately before breakup at $Re$ = 1.2 and $Ca$ = 0.3: (a) planar system, (b) Type 1, and (c) Type 3.}
    \label{fig9}
\end{figure}

To further examine the effect of curvature on the breakup dynamics, \textbf{Fig. ~\ref{fig9}} shows the internal velocity vectors of the drop immediately before breakup for the planar and circular system (Types 1 and 3) at $Re = 1.2$ and $Ca = 0.3$. Since the drop migrates in the circular geometries, the vectors are shown using the velocity relative to the mean velocity inside the drop, $\mathbf{u}_{\mathrm{rel}}=\mathbf{u}_{\mathrm{l}}-\langle\mathbf{u}\rangle_{\mathrm{d}}$. Here, $\mathbf{u}_{\mathrm{l}}$ is the local velocity inside the drop, and $\langle\mathbf{u}\rangle_{\mathrm{d}}$ is the spatially averaged velocity over the drop region. In the planar system, a pair of recirculating motions was formed inside the drop on both sides of the neck \textbf{Fig. ~\ref{fig9} (a)}. This internal flow structure is consistent with the typical shear-induced pinch-off process \citep{Ohtaetal2025}, in which the drop elongates into an inclined prolate shape and then breaks through localized necking near the central region. The recirculating motions on both sides of the neck were relatively balanced, reflecting the nearly symmetric breakup into daughter drops of comparable size. In contrast, the internal flow structure was strongly modified in the circular geometries. In Type 1 shown in \textbf{Fig. ~\ref{fig9} (b)}, a clear recirculating motion was no longer observed in the front part of the migrating drop. Instead, the velocity vectors in this region were mainly directed toward the elongated part of the drop, indicating that the front side was dominated by elongational motion. On the other hand, a localized recirculating motion still remained in the rear part of the drop. This suggests that the curvature first breaks the fore--aft symmetry of the internal circulation: the front part of the drop is stretched in the azimuthal direction, whereas the rear part still retains a recirculating flow structure. This tendency became more pronounced in Type 3 (\textbf{Fig. ~\ref{fig9} (c)}). In this high-curvature case, the internal recirculating motion was strongly suppressed not only in the front part but also in the rear part of the drop. The velocity vectors were mainly aligned along the elongated drop shape, and no clear closed vortex structure was observed inside the thin filament-like region. Therefore, the internal flow in Type 3 was no longer characterized by symmetric recirculation around the neck, but by an elongation-dominated flow along the azimuthal direction. These results indicate that curvature changes the internal flow structure from a recirculation-dominated state, as observed in the planar system, to an elongation-dominated state in the circular geometries. The suppression of internal recirculation, particularly in Type 3, suggests that the drop deformation is no longer governed by localized pinch-off alone. Instead, the drop is stretched over a larger azimuthal extent, leading to the formation and subsequent fragmentation of a long filament-like region.

\begin{figure}[pos=!htbp]
    \centering
    \includegraphics[width=0.85\linewidth]{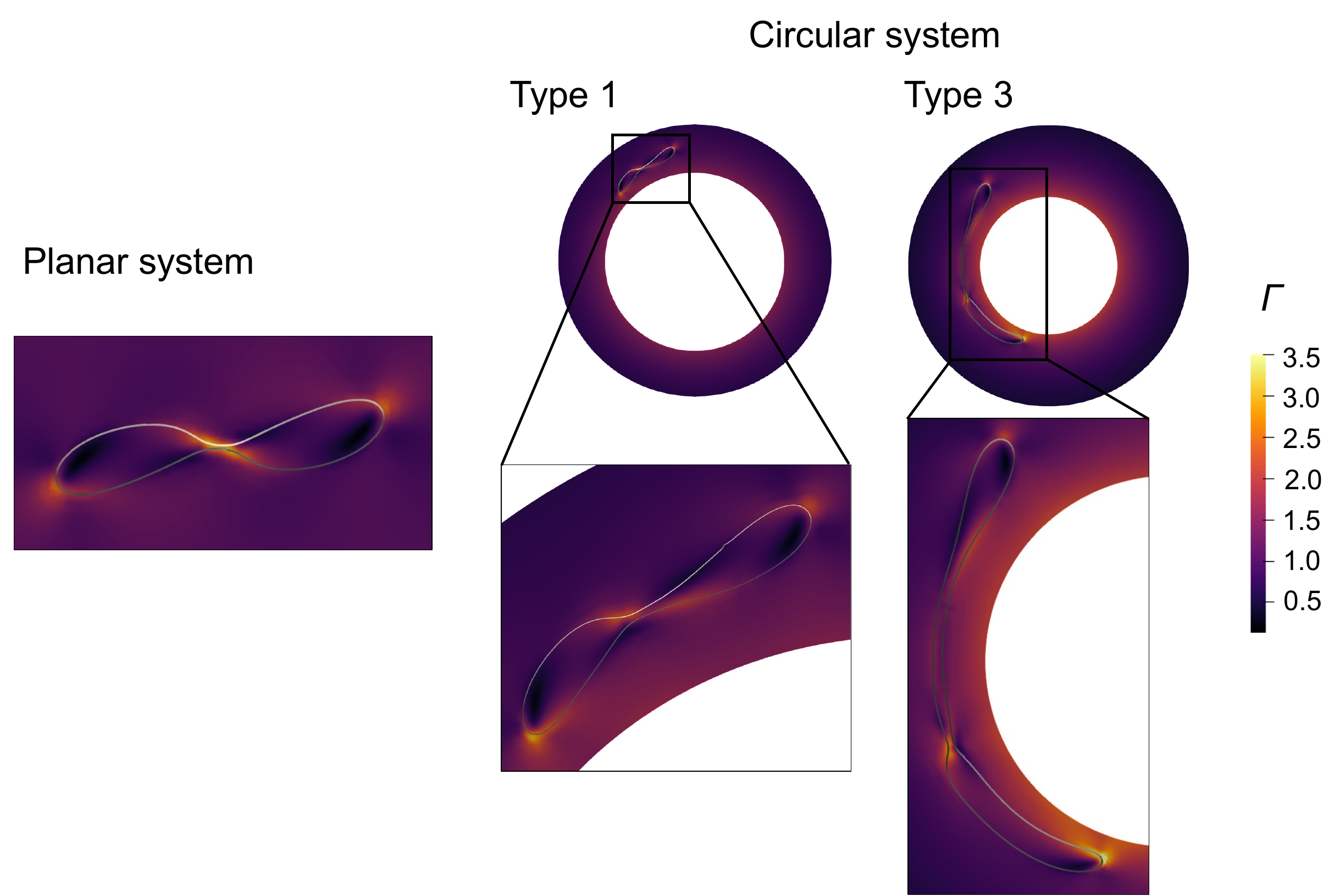}
    \caption{Spatial distribution of dimensionless magnitude of rate-of-deformation tensor, $\varGamma$, immediately before drop breakup at $Re$ = 1.2 and $Ca$ = 0.3. Enlarged views are also provided for circular systems.}
    \label{fig10}
\end{figure}

To clarify how the change in the internal flow structure is related to the local deformation field, the spatial distribution of the magnitude of the rate-of-deformation tensor, $\sqrt{2\mathbf{D}:\mathbf{D}}$, was examined. \textbf{Fig. ~\ref{fig10}} shows the distribution of $\sqrt{2\mathbf{D}:\mathbf{D}}$ normalized by $\dot{\gamma}_{\mathrm R}$, denoted as $\varGamma$, immediately before breakup for the planar and circular systems (Types 1 and 3) at $Re$ = 1.2 and $Ca$ = 0.3. In the planar system, large values of $\varGamma$ were distributed almost symmetrically with respect to the elongated drop axis, and the maximum value appeared near the neck region. This indicates that the local deformation was concentrated around the neck. In addition, as shown in \textbf{Fig. ~\ref{fig9} (a)}, the internal velocity vectors showed a pair of recirculation cells on both sides of the neck, rather than a through-flow directed along the elongated drop. These features suggest that the planar breakup was governed by localized neck thinning, leading to a typical pinch-off process. In the circular geometries, the distribution of $\varGamma$ was markedly different from that in the planar system. In Type 1, large values of $\varGamma$ were observed near the drop tips and the neck region, as in the planar system. However, unlike the planar case, these high-$\varGamma$ regions were no longer distributed symmetrically with respect to the elongated drop axis. In particular, the high-$\varGamma$ region was biased toward the front part of the migrating drop, where the internal velocity vectors in \textbf{Figs. ~\ref{fig9} (b) and (c)} showed no clear recirculation cell. This indicates that the front part of the drop was subjected to local deformation associated with azimuthal elongation, whereas the rear part still retained a localized recirculating flow structure. Therefore, Type 1 can be interpreted as an intermediate case in which the localized pinch-off process observed in the planar system begins to be modified by curvature-induced migration and elongation. This tendency became more pronounced in Type 3. In this high-curvature case, the maximum value of $\varGamma$ appeared near the leading tip of the migrating drop rather than near a central neck region. Moreover, relatively large values of $\varGamma$ were distributed over the elongated filament-like region, instead of being confined to a localized neck. This spatially distributed deformation field agrees with the internal velocity vectors shown in \textbf{Fig. ~\ref{fig9}}. In Type 3, clear recirculation cells were no longer observed inside the elongated drop; instead, the velocity vectors were mainly directed along the filament-like region. Thus, the velocity field and the distribution of $\varGamma$ suggest that azimuthal elongation plays a more prominent role in Type 3 than in the planar system, resulting in the formation of a thin filament that subsequently fragments into multiple satellite drops.

To interpret the differences in drop deformation between the planar and circular systems, we first consider the surrounding flow field in the absence of disturbances caused by the drop. Under this assumption, the shear rate in the circular Couette flow varies radially according to $\dot{\gamma}(r)/\dot{\gamma}_{\mathrm R}=(r_0/r)^2$, whereas the shear rate is spatially uniform in the planar system. Although the reference shear rate at the initial drop center is identical for all geometrical configurations, the variation in the surrounding shear rate across the region initially occupied by the drop becomes more pronounced with increasing geometrical curvature. Furthermore, the initial clearance between the drop and the inner cylinder decreases as the curvature increases. These characteristics may contribute to the asymmetric deformation and enhanced azimuthal elongation observed in the circular systems. Thus, the observed curvature dependence may reflect the combined influences of streamline curvature, shear-rate nonuniformity, and wall proximity.

\begin{figure}[pos=!htbp]
    \centering
    \includegraphics[width=0.5\linewidth]{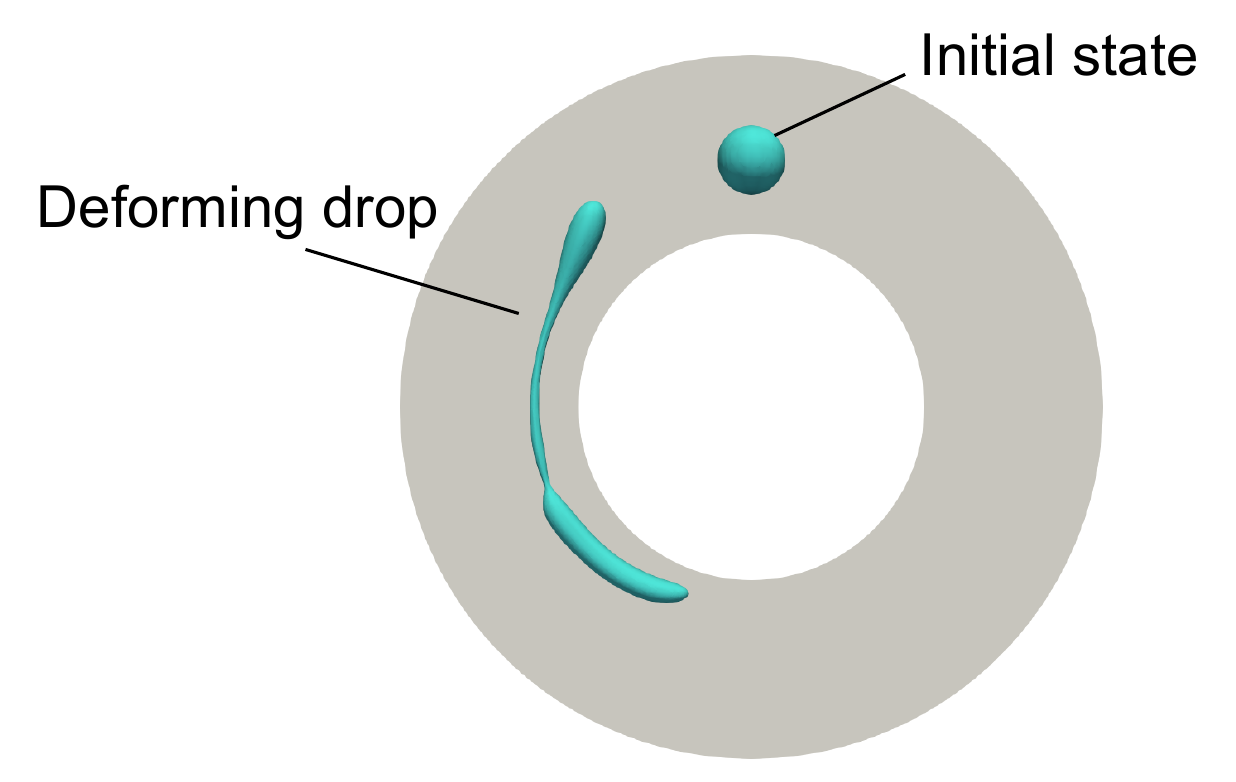}
    \caption{Cross-sectional view of deformed drop at axial midplane for Type 3 at $Re$ = 1.2 and $Ca$ = 0.3.}
    \label{fig11}
\end{figure}

\begin{figure}[pos=!htbp]
    \centering
    \includegraphics[width=0.5\linewidth]{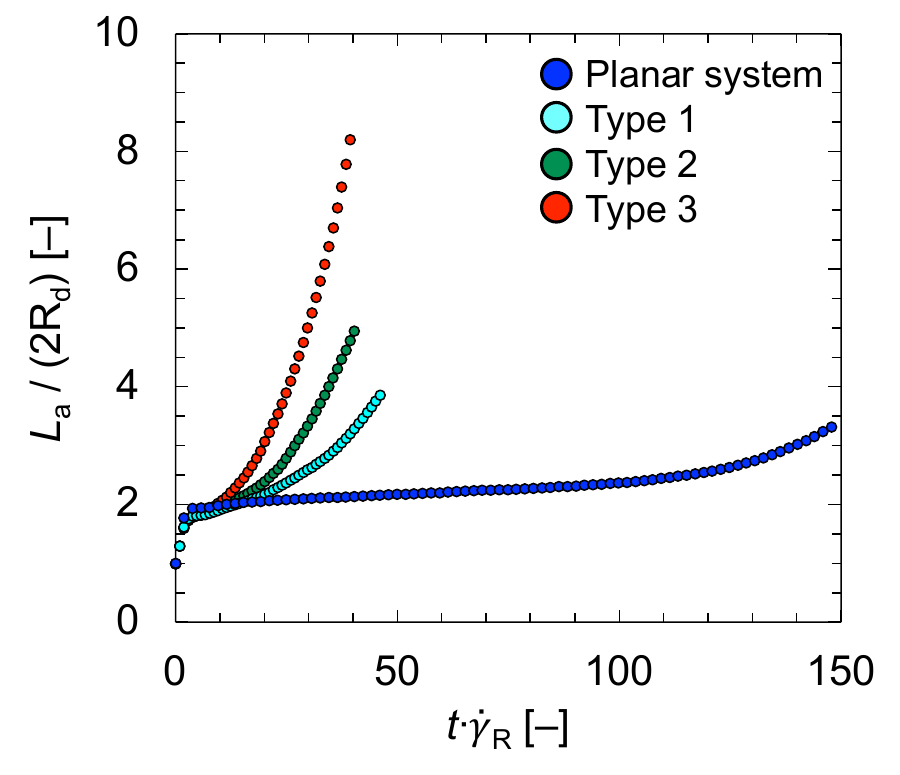}
    \caption{Time evolution of dimensionless elongation length for different geometrical configurations at $Re$ = 1.2 and $Ca$ = 0.3.}
    \label{fig12}
\end{figure}

As indicated in \textbf{Figs. 8--10}, the gap curvature tends to promote drop elongation, particularly in the azimuthal direction. To quantify this geometrical effect on drop deformation and elongation, an appropriate shape parameter is required. As introduced previously, the deformation parameter defined in Eq.~\ref{eq:D} has traditionally been used to characterize drop shape. This parameter, first proposed by \citet{Taylor1934}, represents the drop deformation using characteristic lengths along the major and minor axes. Therefore, it is suitable for nearly ellipsoidal or symmetrically elongated drops, such as those observed in the planar system. However, the highly distorted shape observed in the high-curvature case, particularly for $\kappa_{\mathrm{F}}=1.04$ (\textbf{Fig. ~\ref{fig8}}), deviates significantly from such a simple symmetric deformation and cannot be represented adequately by $D$ alone. Hence, an additional parameter is needed to characterize the azimuthal elongation of the drop. Here, the drop length in the azimuthal direction was selected as a representative parameter. 
\textbf{Fig. ~\ref{fig11}} shows a cross-sectional view of the deforming drop on the mid-plane in the axial direction for Type 3 at $Re = 1.2$ and $Ca = 0.3$, illustrating the definition of the azimuthal elongation length. The angular extent of the drop, $\Delta\theta_a$, was calculated from the reconstructed interface data as the angle between the front and rear tips of the drop along the direction of motion. In the circular geometries, the azimuthal arc length was then defined as $L_{\mathrm a} = r_0\Delta\theta_{\mathrm a}$, where $r_0$ is the initial radial position of the drop. For the planar system, the drop length measured along the major axis was used as the elongation length. \textbf{Fig. ~\ref{fig12}} shows the time evolution of the elongation length for each geometrical configuration up to breakup at $Re = 1.2$ and $Ca = 0.3$. The elongation length and time were nondimensionalized by $2R_{\mathrm d}$ and $\dot{\gamma}^{-1}$, respectively. As clearly shown in \textbf{Fig. ~\ref{fig12}}, the elongation length increased more markedly as the curvature of the annular gap increased, which is consistent with the breakup morphology shown in \textbf{Fig. ~\ref{fig8}}. Furthermore, immediately before breakup, the elongation length increased approximately linearly with time. The slope of this relationship was 0.44 for the planar system, 1.28 for Type 1, 1.96 for Type 2, and 4.8 for Type 3. This result indicates that not only the extent of elongation but also the elongation rate is significantly affected by the gap curvature.

These results demonstrate that, even under the same hydrodynamic non-dimensional parameters, $Re$ and $Ca$, the gap curvature promotes azimuthal longation of the drop and strongly affects the deformation and breakup dynamics.

\subsection{Geometrical effect on critical condition for breakup}\label{subsec:Geometrical effect on critical condition for breakup}
\begin{figure}[pos=!htbp]
    \centering
    \includegraphics[width=0.5\linewidth]{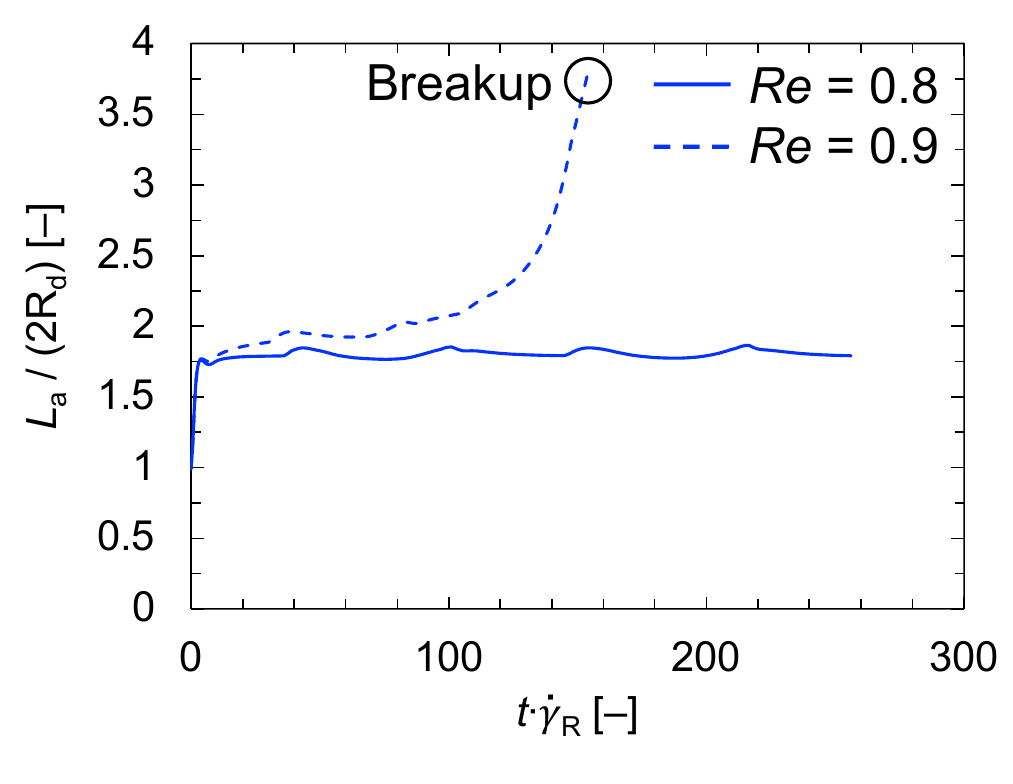}
    \caption{Time evolution of dimensionless elongation length for different geometrical configurations at $Re$ = 1.2 and $Ca$ = 0.3.}
    \label{fig13}
\end{figure}

\begin{figure}[pos=!htbp]
    \centering
    \includegraphics[width=0.85\linewidth]{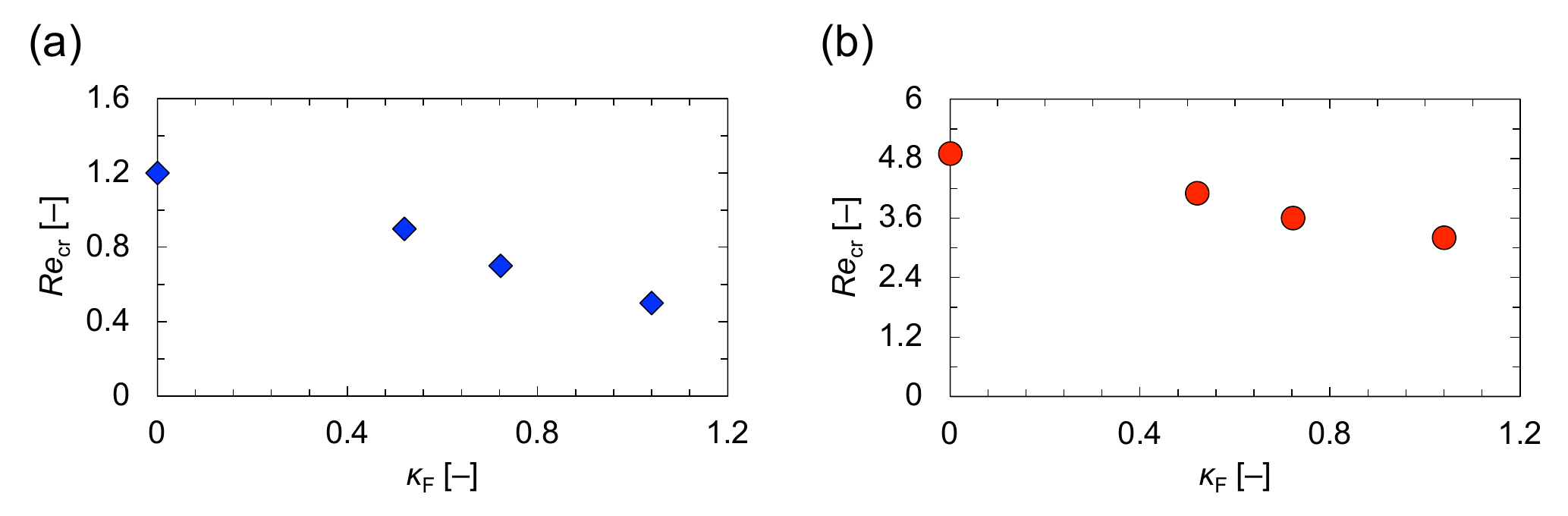}
    \caption{Critical Reynolds number, $Re_{\mathrm{cr}}$, as a function of non-dimensional geometrical curvature, $\kappa_{\mathrm F}$, at (a) $Ca$ = 0.3 and (b) $Ca$ = 0.2.}
    \label{fig14}
\end{figure}

To estimate the critical Reynolds number, simulations were performed by varying $Re$ in increments of 0.1 while keeping $Ca$ constant at either 0.2 or 0.3. For each condition, whether the drop underwent breakup was examined. As an example, \textbf{Fig. ~\ref{fig13}} shows the time evolution of $L_{\mathrm a}/(2R_{\mathrm d})$ for Type 1 at $Re$ = 0.8 and 0.9 with $Ca$ = 0.3. At $Re$ = 0.9, the elongation length increased markedly until the drop broke up into two daughter drops. In contrast, at $Re$ =0.8, the elongation length reached a plateau at approximately 1.8, and no breakup was observed during the simulation. Accordingly, the critical Reynolds number, $Re_{\mathrm{cr}}$, was defined as the lowest $Re$ at which breakup was observed within the investigated range. Thus, $Re_{\mathrm{cr}}$ = 0.9 was obtained for Type 1 at $Ca$ = 0.3. Note that $Re_{\mathrm{cr}}$ was determined with a resolution of 0.1 in the present study. \textbf{Fig. ~\ref{fig14}} compares $Re_{\mathrm{cr}}$ among the different geometrical configurations at $Ca$ = (a) 0.3 and (b) 0.2. The planar geometry corresponds to the limiting case of zero geometrical curvature, i.e., $\kappa_{\mathrm F}=0$, and was therefore plotted at $\kappa_{\mathrm F}=0$. Comparison of \textbf{Figs. ~\ref{fig14}} (a) and (b) shows that $Re_{\mathrm{cr}}$ was higher at $Ca=0.2$ than at $Ca=0.3$ for all geometries. This trend is consistent with the stronger relative influence of surface tension at lower $Ca$, which requires greater inertial effects to induce breakup \citep{Ohtaetal2025}. Furthermore, $Re_{\mathrm{cr}}$ decreased with increasing $\kappa_{\mathrm F}$ at both $Ca$ values, indicating that non-dimensional curvature facilitates drop breakup at lower $Re$. As discussed in Section~\ref{subsec:Geometrical effect on drop dynamics}, this decrease in $Re_{\mathrm{cr}}$ may be associated with the curvature-induced asymmetry of the local flow field around the drop. As $\kappa_{\mathrm F}$ increases, the difference in the local flow field between the inner- and outer-cylinder sides of the drop becomes more pronounced, promoting azimuthal elongation and thereby facilitating breakup at lower $Re$.

\begin{figure}[pos=!htbp]
    \centering
    \includegraphics[width=0.85\linewidth]{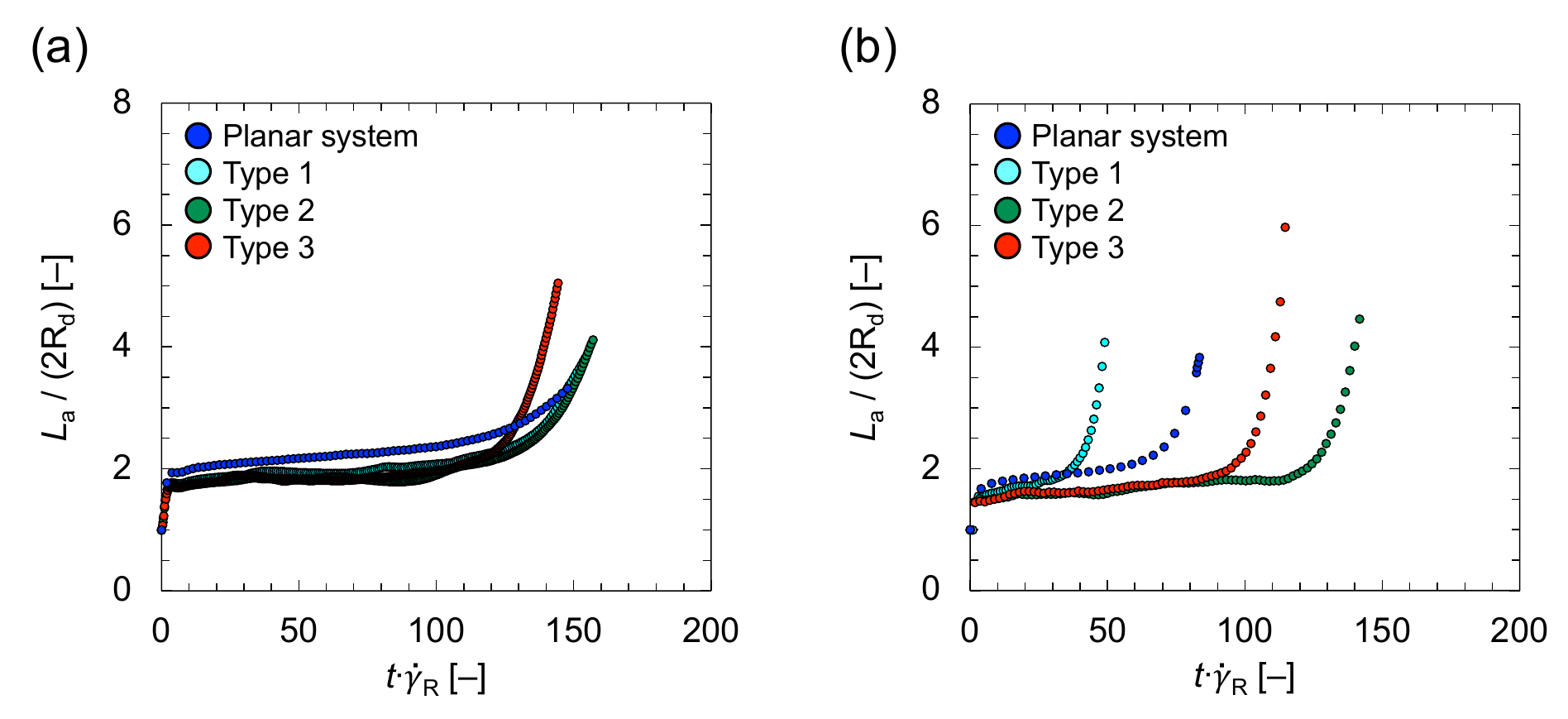}
    \caption{Time evolution of dimensionless elongation length at critical Reynolds number, $Re=Re_{\mathrm{cr}}$, for (a) $Ca$ = 0.3 and (b) $Ca$ = 0.2.}
    \label{fig15}
\end{figure}

\begin{figure}[pos=!htbp]
    \centering
    \includegraphics[width=0.5\linewidth]{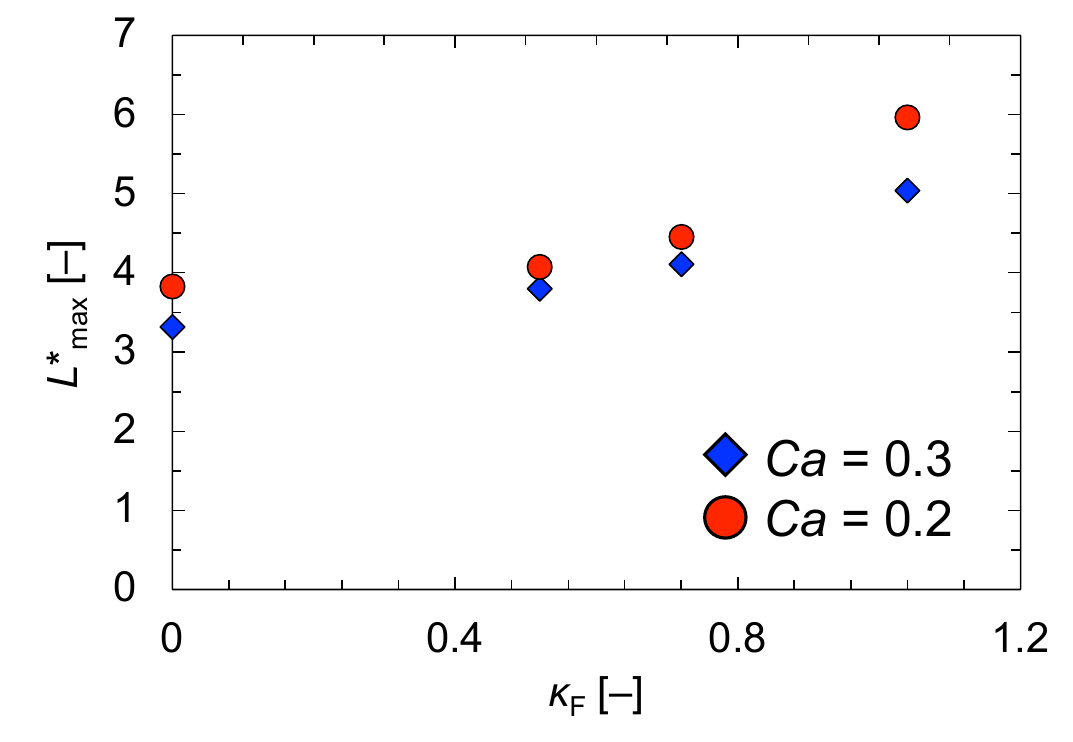}
    \caption{Maximum dimensionless elongation length immediately before breakup as a function of non-dimensional geometrical curvature, $\kappa_{\mathrm F}$.}
    \label{fig16}
\end{figure}

To quantitatively characterize the drop deformation at the critical condition, \textbf{Fig. ~\ref{fig15}} shows the time evolution of the dimensionless elongation length, $L_{\mathrm a}/(2R_{\mathrm d})$, at $Re=Re_{\mathrm{cr}}$ until breakup for $Ca=$ (a) 0.3 and (b) 0.2. Note that the value of $Re_{\mathrm{cr}}$ differs among the geometries and between the two $Ca$ values, as shown in \textbf{Fig. ~\ref{fig14}}. Although no clear dependence of the breakup time on $\kappa_{\mathrm F}$ was observed, the elongation length tended to increase with increasing $\kappa_{\mathrm F}$. To further quantify this trend, \textbf{Fig. ~\ref{fig16}} shows the maximum dimensionless elongation length immediately before breakup, denoted as $L^*_{\mathrm{max}}$, as a function of $\kappa_{\mathrm F}$. For each geometrical configuration, $L^*_{\mathrm{max}}$ was larger at $Ca=0.2$ than at $Ca=0.3$. This difference may be associated with the higher $Re_{\mathrm{cr}}$ required for breakup at lower $Ca$, although the relative influence of surface tension may also affect the deformation process. Moreover, $L^*_{\mathrm{max}}$ increased with $\kappa_{\mathrm F}$ at both $Ca$ values. These results suggest that geometrical curvature affects the extent of drop elongation even at the respective critical conditions for breakup.

As shown in \textbf{Fig. ~\ref{fig12}} in Section~\ref{subsec:Geometrical effect on drop dynamics}, drop elongation became more pronounced with increasing
$\kappa_{\mathrm F}$ at a fixed $Re$ of 1.2. Previous studies have shown that increasing $Re$ promotes drop elongation \citep{Renardy2001,Jeong2026}. However, since $Re_{\mathrm{cr}}$ decreases with increasing non-dimensional curvature, the enhanced drop elongation observed in \textbf{Fig. ~\ref{fig12}} at a fixed $Re$ may simply reflect differences in the relative proximity to the breakup threshold, rather than a change in the breakup mechanism. Nevertheless, as shown in \textbf{Fig. ~\ref{fig16}}, the elongation length immediately before breakup at $Re = Re_{\mathrm{cr}}$ also depends on $\kappa_{\mathrm F}$. This indicates that the difference in drop elongation cannot be explained solely by the reduction in $Re_{\mathrm{cr}}$. Combined with the differences in breakup morphology and internal flow structure discussed in Section~\ref{subsec:Geometrical effect on drop dynamics}, this finding suggests that non-dimensional curvature alters not only the breakup threshold but also the deformation process leading to breakup, even at the respective critical conditions. This finding provides further evidence that non-dimensional curvature influences not only the breakup threshold but also the underlying breakup mechanism. Such effects are particularly relevant to the design of liquid–liquid two-phase flow devices requiring precise control of drop size. Further investigations into the local force balance and the spatiotemporal evolution of interfacial stresses are needed to clarify these mechanisms in greater detail.

\section{Conclusions}\label{sec:Conclusions}
To investigate the effects of geometrical curvature on drop deformation and breakup in shear flow, numerical simulations were performed for a single drop subjected to shear flow between counter-rotating concentric cylinders. The Volume-of-Fluid (VOF) method was employed to capture the interface between the drop and surrounding fluid, using the plicRDF and isoAdvector methods for interface reconstruction and advection, respectively. For simplicity, the density and viscosity were assumed to be identical in both phases. Three circular geometries with different values of the non-dimensional geometrical curvature, $\kappa_{\mathrm F}$, were considered: Type 1 ($\kappa_{\mathrm F}=0.52$), Type 2 ($\kappa_{\mathrm F}=0.72$), and Type 3 ($\kappa_{\mathrm F}=1.04$). The gap width and initial drop radius, were kept constant, while the cylinder radii were varied. The numerical approach was first validated through comparison with published results for drop deformation and breakup in planar shear flow.

First, the effects of geometrical curvature on drop dynamics were examined by comparing the circular and planar systems at fixed Reynolds and Capillary numbers, $Re$ and $Ca$. The results showed that drop elongation became more pronounced with increasing $\kappa_{\mathrm F}$. Furthermore, the breakup morphology differed substantially between the planar and circular systems. In the planar system, breakup occurred through relatively symmetric neck thinning, whereas in the circular systems, the drop exhibited asymmetric deformation and azimuthal elongation. At higher curvature, the drop became highly elongated in the azimuthal direction, forming a thin filament that subsequently broke up into multiple satellite drops.Analysis of the internal velocity field and the magnitude of the rate-of-deformation tensor further revealed differences in the flow and deformation structures associated with breakup. These results suggest that geometrical curvature affects not only the extent of drop elongation but also the deformation process leading to breakup.

Second, the critical Reynolds number, $Re_{\mathrm{cr}}$, was determined by varying $Re$ in increments of 0.1 at fixed $Ca$ values of 0.2 and 0.3. For both $Ca$ values, $Re_{\mathrm{cr}}$ decreased with increasing $\kappa_{\mathrm F}$, indicating that geometrical curvature facilitates drop breakup at lower $Re$. In contrast, the maximum dimensionless elongation length immediately before breakup at $Re=Re_{\mathrm{cr}}$ increased with $\kappa_{\mathrm F}$. This finding indicates that the curvature-induced enhancement of drop elongation cannot be explained solely by the decrease in $Re_{\mathrm{cr}}$. Together with the observed differences in breakup morphology and internal flow structure, these results suggest that geometrical curvature modifies not only the critical condition for breakup but also the deformation process leading to breakup, even at the respective critical conditions.

Future work will examine the effects of cylinder rotation conditions and fluid properties on drop dynamics and the critical conditions for breakup. In the present study, the inner and outer cylinders were rotated in opposite directions with equal magnitudes of tangential wall velocity. Other rotation configurations, including cases in which only one cylinder rotates, will be investigated to assess the influence of the imposed flow field. Furthermore, the effects of density and viscosity ratios between the drop and surrounding fluid remain to be clarified. Detailed investigations of the local force balance and interfacial dynamics will also be necessary to elucidate the underlying breakup mechanisms in curved shear flows.

\printcredits

\section*{Acknowledgements}
This research was partially supported by JSPS KAKENHI (grant numbers 21KK0261, 23K13592, and 25K08368). Calculations for this research were conducted on the Lichtenberg high performance computer of Technische Universität Darmstadt. The authors would like to thank Dr. Jun Liu and Mr. Muhammad Hassan Asghar (Technische Universität Darmstadt) for their valuable technical support in conducting the numerical simulations and developing the computational code.

\section*{Declaration of generative AI and AI-assisted technologies in the writing process}
During the preparation of this work, the authors used ChatGPT-5.5 in order to improve readability and language. After using this tool, the authors reviewed and edited the content as needed and take full responsibility for the content of the publication.

\section*{Declaration of competing interest}
The authors have declared no conflict of interest.

\nomenclature[A]{$Ca$}{Capillary number}
\nomenclature[A]{$Re$}{Reynolds number}
\nomenclature[A]{$Re_{\mathrm {cr}}$}{Critical Reynolds number}
\nomenclature[A]{$d$}{Gap width (m)}
\nomenclature[A]{$D$}{Deformation parameter (m)}
\nomenclature[A]{$H$}{Axial height of domain (m)}
\nomenclature[A]{$L_a$}{Azimuthal arc length (m)}
\nomenclature[A]{$L^*_{\mathrm{max}}$}{Maximum dimensionless elongation length}
\nomenclature[A]{$R_{\mathrm d}$}{Initial drop radius (m)}
\nomenclature[A]{$R_{\mathrm i}$}{Inner cylinder radius (m)}
\nomenclature[A]{$R_{\mathrm o}$}{Outer cylinder radius radius (m)}

\nomenclature[A]{$r$}{Initial position of drop in radial direction (m)}
\nomenclature[A]{$\mathbf D$}{Rate-of-deformation tensor (s$^{-1}$)}
\nomenclature[A]{$L_{\mathrm d}$}{Major axis of deformed drop (m)}
\nomenclature[A]{$B_{\mathrm d}$}{Minor axis of deformed drop (m)}

\nomenclature[A]{$t$}{Time (s)}
\nomenclature[A]{$p$}{Pressure (Pa)}
\nomenclature[A]{$\mathbf g$}{Gravitational acceleration (m $\cdot$ s$^{-2}$}
\nomenclature[A]{$U$}{Wall velocity (m $\cdot$ s$^{-1}$))}
\nomenclature[A]{$\mathbf u$}{Velocity (m $\cdot$ s$^{-1}$))}
\nomenclature[A]{$\mathbf u_{\mathrm {rel}}$}{Relative velocity (m $\cdot$ s$^{-1}$))}
\nomenclature[A]{$\mathbf u_{\mathrm {l}}$}{Local velocity (m $\cdot$ s$^{-1}$))}
\nomenclature[A]{$\langle\mathbf{u}\rangle_{\mathrm{d}}$}{Spatially averaged velocity over drop region (m $\cdot$ s$^{-1}$))}
\nomenclature[A]{$\mathbf n_{\Sigma}$}{Normal unit vector at interface}

\nomenclature[A]{$u_{\theta}$}{Azimuthal velocity (m $\cdot$ s$^{-1}$))}
\nomenclature[A]{$V$}{Volume [m$^3$]}

\nomenclature[A]{$\ell$}{Mesh refinement level}

\nomenclature[G]{$\Delta x, \Delta y, \Delta z$}{Cell sizes in the $x$, $y$, and $z$ directions [m]}
\nomenclature[G]{$\alpha_{\mathrm c}$}{Cell-averaged volume fraction of drop phase}
\nomenclature[G]{$\chi$}{Phase indicator function}
\nomenclature[G]{$\delta_{\Sigma}$}{Dirac delta distribution}
\nomenclature[G]{$\Gamma$}{Dimensionless deformation rate}
\nomenclature[G]{$\kappa_{\mathrm F}$}{Geometrical curvature}
\nomenclature[G]{$\dot{\gamma}$}{shear-rate (s$^{-1}$)}
\nomenclature[G]{$\dot{\gamma}_{\mathrm R}$}{Representative shear-rate (s$^{-1}$)}
\nomenclature[G]{$\Delta\theta_a$}{Angular extent of drop (rad)}
\nomenclature[G]{$\sigma$}{Interfacial tension (N $\cdot$ m$^{-1}$)}
\nomenclature[G]{$\kappa$}{Interfacial curvature (m$^{-1}$)}
\nomenclature[G]{$\rho$}{Fluid density (kg $\cdot$ m$^{-3}$)}
\nomenclature[G]{$\eta$}{Fluid viscosity (Pa $\cdot$ s)}

\nomenclature[G]{$\Omega$}{Computational domain}
\nomenclature[G]{$\Omega^+$}{Domain occupied by the drop}
\nomenclature[G]{$\Omega^-$}{Domain occupied by the surrounding fluid}
\nomenclature[G]{$\Omega_{\mathrm c}$}{Control volume}
\nomenclature[G]{$\Sigma$}{Fluid--fluid interface}

\nomenclature[G]{$\varGamma$}{Dimensionless deformation rate}
\nomenclature[G]{$\kappa_{\mathrm F}$}{non-dimensional geometrical curvature}
\nomenclature[G]{$\dot{\gamma}$}{shear-rate (s$^{-1}$)}
\nomenclature[G]{$\omega_{\mathrm i}$}{Angular velocity of inner cylinder (rad $\cdot$ s$^{-1}$)}
\nomenclature[G]{$\omega_{\mathrm o}$}{Angular velocity of outer cylinder (rad $\cdot$ s$^{-1}$)}
\nomenclature[S]{$\mathrm d$}{Drop}
\nomenclature[S]{$\mathrm s$}{Surrounding fluid}

\printnomenclature






\clearpage 




\renewcommand{\bibsection}{\section*{References}}
\bibliographystyle{cas-model2-names}

\bibliography{cas-refs}



\end{document}